\documentclass[titlepage,12pt]{article}
\usepackage{textcomp}
\usepackage{chetnew}
\usepackage{amssymb,amsmath,color,graphics,amscd,epsf,indentfirst,amsfonts}
\usepackage{mathtools}
\usepackage{enumerate}
\usepackage{epsfig}
\usepackage{slashed}

\usepackage{tikz}

\def\blfootnote{\xdef\@thefnmark{}\@footnotetext}

\long\def\symbolfootnote[#1]#2{\begingroup%
\def\thefootnote{\fnsymbol{footnote}}\footnote[#1]{#2}\endgroup}

\makeatletter
\renewcommand{\@dotsep}{4.5}
\makeatother

\def\be{\begin{equation}}
\def\ee{\end{equation}}

\makeatletter
\def\@seccntformat#1{\csname the#1\endcsname.\quad}
\makeatother

\def\clock{{\count0=\time
           \divide\count0 60
           \ifnum\count0<10 0\fi\the\count0
           \multiply\count0 -60 \advance\count0 \time
           :\ifnum\count0<10 0\fi \the\count0
         }}
\newcommand{\timestamp}{{\small\vbox{\hbox{\tt\jobname.tex}
\hbox{\the\day/\the\month/\the\year, \clock}}}}

\def\time{{\tau}}

\def\beq{\begin{equation}}
\def\eeq{\end{equation}}
\newcommand{\bea}{\begin{eqnarray}}
\newcommand{\eea}{\end{eqnarray}}
\def\bal{\begin{align}}
\def\eal{\end{align}}

\def\drawbox#1#2{\hrule height#2pt
         \hbox{\vrule width#2pt height#1pt \kern#1pt
               \vrule width#2pt}
               \hrule height#2pt}

\def\Asym#1#2{\vcenter{\vbox{\drawbox{#1}{#2}
               \kern-#2pt       
               \drawbox{#1}{#2}}}}

\numberwithin{equation}{section}

\begin{document}
\begin{titlepage} 
\vskip 4cm
\begin{center}
\font\titlerm=cmr10 scaled\magstep4
    \font\titlei=cmmi10 scaled\magstep4
    \font\titleis=cmmi7 scaled\magstep4
    \centerline{\LARGE \titlerm 
      Dualities of Adjoint QCD$_3$ from Branes}
          \vskip 0.3cm
\vskip 1cm
{Adi Armoni}\\
\vskip 0.5cm
       {\it Department of Physics, Faculty of Science and Engineering}\\
       {\it Swansea University, SA2 8PP, UK}\\
\medskip
\vskip 0.5cm
{a.armoni@swansea.ac.uk}\\

\end{center}
\vskip .5cm
\centerline{\bf Abstract}

\baselineskip 20pt
%

\vskip .5cm 
\noindent
We consider an 'electric' $U(N)$ level $k$ QCD$_3$ theory with one adjoint Majorana fermion. Inspired by brane dynamics, we suggest that for $k \ge N/2$ the massive $m<0$ theory, in the vicinity of the supersymmetric point, admits a  $U(k-\frac{N}{2})_{-(\frac{1}{2}k+\frac{3}{4}N),-(k+\frac{N}{2})}$ 'magnetic' dual with one adjoint Majorana fermion. The magnetic theory flows in the IR to a topological $U(k-\frac{N}{2})_{-N,-(k+\frac{N}{2})}$ pure Chern-Simons theory in agreement with the dynamics of the electric theory. When $k<N/2$ the magnetic dual is $U(\frac{N}{2}-k)_{\frac{1}{2}k+\frac{3}{4}N,N}$ with one adjoint Majorana fermion. Depending on the sign of the fermion mass, the magnetic theory flows to either $U(\frac{N}{2}-k)_{N,N}$ or $U(\frac{N}{2}-k)_{\frac{1}{2}N+k,N}$ TQFT. A second magnetic theory,  $U(N/2+k)_{\frac{1}{2}k-\frac{3}{4}N,-N}$, flows to either  $U(\frac{N}{2}+k)_{-N,-N}$ or $U(\frac{N}{2}+k)_{-(\frac{1}{2}N-k),-N}$ TQFT. Dualities for $SO$ and $USp$ theories with one adjoint fermion are also discussed.

\vfill
\noindent
\end{titlepage}\vfill\eject

\hypersetup{pageanchor=true}

\setcounter{equation}{0}

\pagestyle{empty}
\small
\vspace*{-0.7cm}
{
\hypersetup{linkcolor=black}
\tableofcontents
}
\normalsize
\pagestyle{plain}
\setcounter{page}{1}

\section{Introduction} 
\label{intro}

Quantum chromodynamics in three spacetime dimensions (QCD$_3$) attracted in recent years a lot of attention (see \cite{Komargodski:2017keh} and references within). The vacuum structure of the theory is rich as it depends on the number of colours, the number of flavours, the Chern-Simons (CS) level, the representation of the quarks and their masses. 

One particular interesting variant of the theory is QCD with a single Majorana adjoint fermion. When the mass is large it can be integrated out, resulting in a shift of the Chern-Simons level as follows
\beq
 k \rightarrow k + {\rm sgn}(m) \frac{1}{2} h(G) \, , \label{shift}
 \eeq
 where $h(G)$ is the dual Coexter number of the group. Due to the level shift it follows that the theory admits at least one phases transition as the mass of the fermion is varied. The phases of the theory were recently studied in \cite{Gomis:2017ixy}, where it was argued that for $k \ge \frac{1}{2} h(G)$ the theory admits two phases, while surprisingly for  $k < \frac{1}{2} h(G)$ the theory admits three phases: apart from the large positive mass phase and the large negative mass phase, it admits an intermediate quantum phase. The physics in the vicinity of the two phase transitions is described by two dual theories. Follow-up papers which are relevant to the current work include the generalisations to two adjoint flavours \cite{Choi:2019eyl} and to other two-index representations \cite{Choi:2018tuh}.
 
 One of the striking results of \cite{Gomis:2017ixy} is that $SU(N)_0$\footnote{Throughout this paper we use the notation $G_k$ for a level $k$ Chern-Simons theory based on gauge group $G$.} with one adjoint fermion is not a confining theory. Instead, the IR of the theory is described by $U(\frac{N}{2})_{\frac{N}{2},N}$ TQFT. As a result the Wilson loop in the fundamental representation shows a perimeter law. A convincing argument in favour of this result is anomaly matching: the UV theory cannot flow to a trivial theory in order to match the anomalous $1$-form symmetry, see \cite{Lohitsiri:2022jyz} for a recent discussion.
 
In this note we use branes to re-derive the dualities of \cite{Gomis:2017ixy} as well as new dualities. We will use a 3d version of Seiberg duality, closely related to the Giveon-Kutasov duality  \cite{Giveon:2008zn}. The electric theory is $U(N)_k$\footnote{Here we specify the level of the $SU(N)$ part. The $U(1)$ level is rather subtle and will be discussed later.} (or $USp(2N)_{2k}$, $SO(2N)_{2k}$, $SO(2N+1)_{2k}$) with one adjoint fermion. The Seiberg dual for $k \ge \frac{1}{2} h(G)$ is supersymmetric for a certain value of the fermion mass. The duality is valid in the vicinity of this value. As we shall see both electric and magnetic theories admit the same (Witten) index. Moreover, they both flow to the same IR theory.

When $k < \frac{1}{2} h(G)$ (in analogy with $N_f<N_c$ in 4d SQCD) supersymmetry is broken. Yet, we are able to recover {\it two magnetic} duals whose matter content was conjectured in \cite{Gomis:2017ixy}. 

Therefore, string theory beautifully recovers known dualities for $k<\frac{1}{2}h(G)$ and, moreover, predicts new dualities for $k\ge \frac{1}{2}h(G)$ that can be verified by using several consistency checks such as anomaly matching, agreement of an index and the coincidence of the low energy dynamics at the weak coupling regime of large fermion mass.

Throughout the paper we use $h(U(N))=N$, $h(USp(2N))=2N+2$, $h(SO(2N))=2N-2$. We use the convention $USp(2)\sim SU(2)$.

\section{Brane dynamics}
\label{duality}

\begin{figure}[!th]
\centerline{\includegraphics[width=8cm]{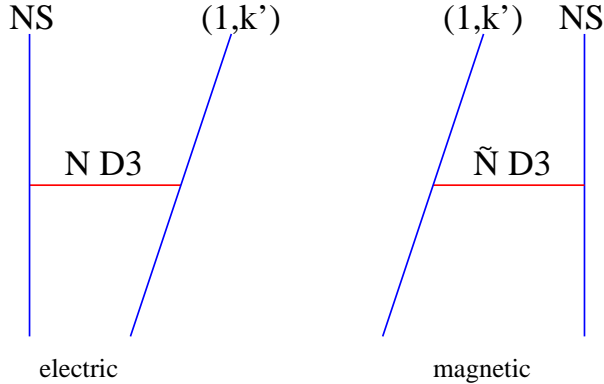}}
\caption{\footnotesize Seiberg Duality from branes. The electric theory (left) is a $U(N)_{k',k'}$ gauge theory with one massive $m^2 \gg g^2$ adjoint fermion and an additional light adjoint fermion. Similarly, the magnetic theory (right) is $U(\tilde N)_{-k',-k'}$ gauge theory with one massive $m^2 \gg g^2$ adjoint fermion and an additional light adjoint fermion. $k'=k+N/2$.}
\label{branes}
\end{figure}

We derive the gauge theory dualities from a duality in string theory. Our setup is similar to that of refs.\cite{Armoni:2014cia} and \cite{Armoni:2017jkl}. We consider 'electric' and 'magnetic' branes configurations as depicted in figure \eqref{branes}. The configurations consist of $N$ (or $\tilde N$) D3 branes suspended between NS fivebrane and a tilted $(1,k')$ fivebrane. The D3 branes span the $012$ directions and a finite segment of the $6$ direction. The NS fivebrane spans the $012345$ directions. The $(1,k')$ fivebrane spans the $01238$ directions and it is tilted in the $(59)$ plane with an angle $-\frac{\pi}{2}-\theta$. In order to preserve supersymmetry we choose $\tan \theta = g_s k'$. The content of the theory admits 3d ${\cal N}=2$ supersymmetry and it can be obtained by dimensional reduction of 4d ${\cal N}=1$ SYM. The CS interaction reduces the supersymmetrey to 3d ${\cal N}=1$. The Lagrangian of the theory may be written in terms of ${\cal N}=1$ supersymmetry as a level $k'$ vector multiplet $V=(A_\mu, \chi)$ coupled to an adjoint scalar multiplet $\Phi = (\phi,\psi)$. The action is given by\footnote{The same brane configuration was used in \cite{Armoni:2009vv} to construct the Acharya-Vafa theory \cite{Acharya:2001dz} of ${\cal N}=1$ SYM domain walls.}
\bea
& & S=\frac{1}{4g^2} \int \text{d}^3x ~
{\rm tr} \left( (D \phi)^2-F^2+i \bar \chi \displaystyle{\not}D \chi
+i \bar \psi \displaystyle{\not}D \psi+2i \bar \chi[\phi,\psi]
\right)+ \\ \nonumber
& & + \frac{k'}{4\pi} \int {\rm tr} \left( A\wedge \text{d}A+\tfrac{2}{3} A\wedge A\wedge A \right)
-\frac{k'}{4\pi} \int \text{d}^3 x~\bar \chi \chi~. \nonumber
\eea
Let us begin with the electric theory. In order to obtain a low energy theory with a gauge group $U(N)$ and a light adjoint fermion $\chi$, we choose $k'=k+N/2$. We then give a large negative mass to $\psi$. A mass term for the fermion that lives on the branes can be achieved by tilting the $(1,k')$ fivebrane further, or by turning on background fluxes \cite{Camara:2003ku}. In particular in ref.\cite{Camara:2003ku} the authors discussed in detail how a 3-form background flux in type IIB can be used to give mass to matter that lives on D3 branes.

The level of the $SU(N)$ part is shifted by $-N/2$ due to the integration over the massive adjoint fermion, $\psi$.\footnote{It is also possible to begin with $-k'=-k+N/2$ and to give a  positive mass to $\psi$. When we do so, the resulting dual magnetic theory covers a different range of masses of the electric theory. We will later refer to that theory as magnetic'.} Note that the $U(1)$ level is not shifted because the adjoint fermion does not carry a $U(1)$ charge. The resulting electric theory is a 3d ${\cal N}=1$ $U(N)_{k,k'}$ YM-CS theory. 

We follow the prescription of Elitzur-Giveon-Kutasov \cite{Elitzur:1997fh} and swap the fivebranes in order to obtain the Seiberg dual, see fig.\eqref{branes}. We start with $k\ge N/2$ in order to preserve supersymmetry. The resulting magnetic dual is a ${\cal N}=1$ supersymmetric $U(\tilde N)_{-\tilde k,-k'}$ YM-CS theory\footnote{The levels of the magnetic theory are negative because the fundamental string ends on the left fivebrane, as opposed to the right fivebrane in the electric theory case. It is as if we integrated out an infinitely heavy massive flavour, with a negative mass.}. As we swap the branes the number of D3 branes in the magnetic side becomes (similar to $N_f-N_c$ D4 branes in 4d SQCD \cite{Elitzur:1997fh})
\beq
\tilde N=k'-N=k-\frac{1}{2}N \, .
\eeq
The level of the $SU(\tilde N)$ part in the magnetic theory, $\tilde k$, is given by $k'$ shifted due to the massive fermion in the scalar multiplet
\beq
\tilde k = k' - \frac{1}{2} \tilde N = k + \frac{1}{2}N - \frac{1}{2} (k-\frac{1}{2}N) = \frac{1}{2} k + \frac{3}{4} N \,  .
\eeq

We therefore propose the following Seiberg duality between two ${\cal N}=1$ supersymmetric YM-CS theories:
\beq
U(N)_{k,k+\frac{N}{2}} \iff U(k-N/2)_{-(\frac{1}{2}k+\frac{3}{4}N),-(k+\frac{N}{2})} \label{new-duality} \,.
\eeq
The above duality \eqref{new-duality} is the main result of this note.

The first check of the duality \eqref{new-duality} is the Witten index. The electric  ${\cal N}=1$ $U(N)_{k,k+\frac{N}{2}}$ theory as well as the $U(k-N/2)_{-(\frac{1}{2}k+\frac{3}{4}N),-(k+\frac{N}{2})}$ admit the same Witten index \cite{Witten:1999ds,Choi:2018ohn,Bashmakov:2018ghn,Delmastro:2020dkz}
\beq
I_W=\frac{(k+\tfrac{1}{2}N)!}{(k-\tfrac{1}{2}N)!N!} \label{witten-index} \, .
\eeq

The Witten index can be derived from the brane configuration by using the s-rule and counting all possible ways the $N$ (or $\tilde N$) D3 branes can end on the $k+N/2$ constituent fivebranes \cite{Bergman:1999na,Ohta:1999iv}. Both the electric and magnetic descriptions lead to \eqref{witten-index}, in agreement with field theory.

Let us briefly discuss the case with $k<N/2$. When we swap the fivebranes we obtain $\tilde N=k-\frac{1}{2}N$ {\it anti} D3 branes suspended between the fivebranes. The brane configuration breaks supersymmetry, in agreement with the expectation from the gauge dynamics of the electric theory when $k<N/2$. Typically a non-supersymmetric brane configuration is unstable and cannot be trusted, however in the current configuration the D3 branes are locked between the fivebranes and therefore the 3d field theory does not suffer from an instability. An analogoues scenario in 4d SQCD leads to $N_c-N_f$ antibranes suspended between orthogonal NS fivebranes, however the situation in 4d is more complex because there are mesons and thus one may explain the Affleck-Dine-Seiberg runaway superpotential by a potential between branes.\footnote{This argument is due to Shigeki Sugimoto.}
The 3d field theory that lives on the branes is a non-supersymmetric $U(N/2-k)_{\frac{1}{2}k+\frac{3}{4}N}$ YM-CS theory with a Majorana adjoint fermion. Supersymmetry breaking occurs due to the interaction between the {\it anti} D3 branes and the fivebranes.  Our results in section \eqref{k<N/2} for $k<N/2$ are in full agreement with the dualities proposed in \cite{Gomis:2017ixy} and reveals their string theory origin.  

Our picture and the conjectured dualities are supported by the gravity dual, proposed in \cite{Maldacena:2001pb}. The dual of $SU(N)_k$ with ${\cal N}=1$ supersymmetry is described by a confining geometry when $k=N/2$. When $k$ is increased additional branes are added to the geometry in a SUSY preserving manner. When $k$ is decreased {\it antibranes} are added and supersymmetry is broken. The emerging picture from our setup is essentially the same. The advantage of the current dualities over the holographic dual is that it applies to finite $N$ as well as for a wide range of masses, hence it is useful for exploring phase transitions. 

In the following sections we will study the field theory dynamics of the electric and magnetic theories in more detail.

\section{Gauge dynamics when $k \ge N/2$}
\label{k>N/2}

The electric $U(N)_{k,k+N/2}$ theory is supersymmetric when the bare fermion mass is $m=-g^2k$. It is conjectured that the theory admits two phases. The low energy dynamics is characterised by two TQFT's. At large positive mass it is a $U(N)_{k+N/2,k+N/2}$ TQFT and at large negative mass it is a $U(N)_{k-N/2,k+N/2}$ TQFT. 

\vskip 0.5cm
\begin{figure}[!th]
\centerline{\includegraphics[width=14cm]{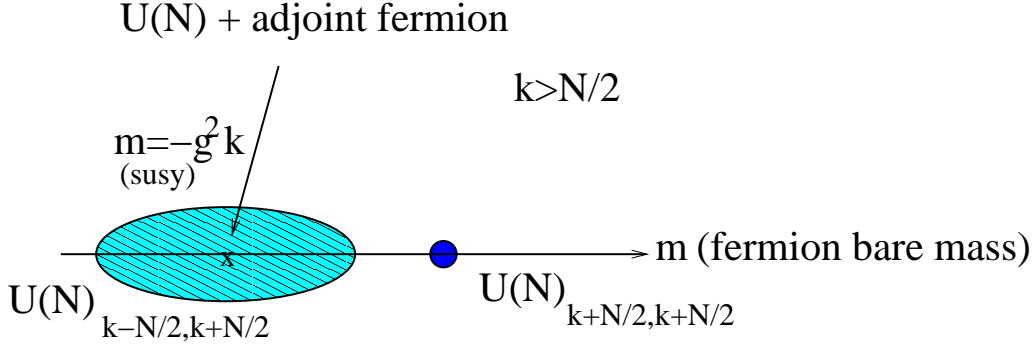}}
\caption{\footnotesize The phase diagram of adjoint QCD with $k \ge N/2$. There are two phases, separated by a phase transition denoted by the blue dot. The supersymmetric point lies at $m=-g^2k$. The dual gauge theory (shaded ellipse) describes the phase with negative mass in the vicinity of the supersymmetric point. When $k=N/2$ the $SU(N)$ IR theory is trivial (gapped).}
\label{phase1}
\end{figure}
\vskip 0.3cm

The proposed magnetic dual describes the supersymmetric point and its vicinity. Starting with $U(k-N/2)_{-(\frac{1}{2}k+\frac{3}{4}N),-(k+N/2)}$ let us integrate out the massive adjont fermion. The resulting theory is $U(k-N/2)_{-N,-(k+N/2)}$ which is dual to $U(N)_{k-N/2,k+N/2}$ \cite{Hsin:2016blu}. We obtained the IR TQFT in the vicinity of the supersymmetric electric theory by using the dual magnetic theory, see fig.\eqref{phase1}. In addition, at the supersymmetric point there is an additional adjoint fermion, as expected from the electric side. As we already discussed, the Witten index of the dual pair matches. 

At $k=N/2$ the dual magnetic theory is trivial, apart from a free massless $U(1)$. This is not surprising: the $SU(N)$ part of the electric theory is expected to confine and hence the IR is gapped, except a free $U(1)$.

\section{Gauge dynamics when $k<N/2$}
\label{k<N/2}

\begin{figure}[!th]
\centerline{\includegraphics[width=14cm]{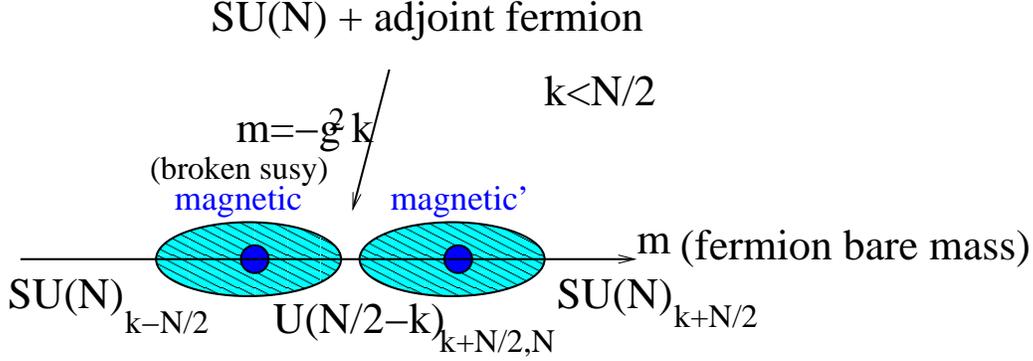}}
\caption{\footnotesize The phase diagram of adjoint QCD with $k < N/2$. There are three phases, separated by two phase transitions denoted by blue dots. The supersymmetric point lies at $m=-g^2k$. The magnetic theory (shaded ellipse) describes the vicinity of the (broken-)supersymmetric point, at masses both below and above the phase transition. The magnetic' theory covers regions above and below the positive mass phase transition.}
\label{phase2}
\end{figure}

When $k<N/2$ the electric theory admits three phases. In addition to the large positive mass and large negative masses, there is an intermediate quantum phase characterised by a TQFT \cite{Gomis:2017ixy}, see fig.\eqref{phase2}
\beq
 U(N/2-k)_{N/2+k,N} \sim  U(N/2+k)_{-N/2+k,-N} \,.
 \eeq
 Within this phase there exists a point with a massless Goldstino, due to the spontaneous breaking of supersymmetry.

\vskip 0.5cm
\vskip 0.5cm

Naively, the proposed dual YM-CS is $U(N/2-k)_{\frac{1}{2}k+\frac{3}{4}N,k+N/2}$ with an adjoint fermion. However, such naive assignment of the $U(1)$ level is not compatible with the dynamics of the electric theory. Since we have two magnetic duals we need to make sure that their $U(1)$ levels reproduce correctly the intermediate phase. In order to have a consistent picture we need to assign a level $N$ to the $U(1)$ of the magnetic theory (and level $-N$ to the magnetic' theory). The shift of the $U(1)$ level from its naive value of $k+N/2$ to the `correct' $N$ may be achieved by introducing a RR 0-form flux along the $6$ direction, or perhaps a localised flux at the intersection of the D3 antibranes and the fivebrane
\beq 
 S=\int dx^6\, \int d^3x\, C_0 F \wedge F = -\int dx^ 6\, dC_0 \int d^3 x \, A \wedge F  \, .
 \eeq
This is a subtle matter and a better understanding of non-supersymmetric brane configuration which include antibranes is needed.

Let us integrate out the adjoint fermion. Positive and negative masses results in two TQFT. The negative mass results in $U(N/2-k)_{(k+N/2),N}$ TQFT and the positive mass results in $U(N/2-k)_{N,N}$ (which is dual to $SU(N)_{k-N/2,-N}$). Thus the dual gauge theory describes two regimes of the three phases of the electric theory: the negative mass phase and the intermediate phase, see fig.\eqref{phase2}. In addition at some value of the mass there is a massless fermion which corresponds to the Goldstino of the electric theory. Similar to the magnetic theory, there exists a magnetic' theory that covers a regime of the positive mass phase and the intermediate mass phase. The details are as follows: the magnetic' theory is $U(N/2+k)_{\frac{1}{2}k-\frac{3}{4}N,-N}$ with an adjoint fermion. After integrating out the massive fermion we end up with either $U(N/2+k)_{-(N/2-k),-N}$ TQFT or $U(k+\frac{1}{2}N)_{-N,-N}$ TQFT. They describe (by virtue of level-rank duality) the  $U(N/2-k)_{N/2+k,N}$  and $SU(N)_{k+\frac{1}{2}N}$ phases of the electric theory, see fig.\eqref{phase2}.

Our results in this section are in agreement with ref.\cite{Gomis:2017ixy}.

\section{SO and USp theories}
\label{SOSp}

The generalization to SO and USp theories is straightforward, yet we will encounter new phenomena. We use the same brane configuration that led to the $U(N)$ dualities, with minor modifications. 

 The USp theory is realized  by adding to the brane configuration an $O3^+$ plane along the 0126 directions. The orientifold plane does not tolerate a fractional brane, therefore we need to place $N$ D3 branes and their mirrors, namely an even number of branes. The resulting theory is $USp(2N)$. Similarly, $SO(2N)$ and $SO(2N+1)$ theories are realized by placing either even or odd number of D3 branes on top of an $O3^-$ plane. In both USp/So cases we will use a $(1,2k')$ tilted fivebrane.

\subsection{The $USp(2N)$ theory}
 We begin with the electric theory. Let $k'=k+ \frac{1}{2}(N+1)$. The adjoint fermion with large negative mass shifts the level such that the resulting electric theory that lives on the brane configuration is $USp(2N)_{2k}$.
  
 As we swap the fivebranes an additional D3 antibrane (and its mirror)  is created between the fivebranes due to the presesnce of the orientifold plane. Therefore $\tilde N = k'-1-N=k-\frac{1}{2}(N+1)$ and $\tilde k = k'-\frac{1}{2}(\tilde N +1)=\frac{1}{2}k + \frac{3}{4}N + \frac{1}{4}$.
The resulting magnetic theory is ${\cal N}=1$ supersymmetric $USp(2(k-\frac{1}{2}N-\frac{1}{2}))_{-2(\frac{1}{2}k +\frac{3}{4}N + \frac{1}{4})}$ YM-CS theory, hence we propose the following Seiberg duality
\beq
 USp(2N)_{2k} \iff USp(2(k-\frac{1}{2}N-\frac{1}{2}))_{-2(\frac{1}{2}k +\frac{3}{4}N + \frac{1}{4})}
\eeq
 The supersymmetric magnetic dual can be trusted when $\tilde N = k-\frac{1}{2}(N+1) \ge 0$. By using the assignment $N\rightarrow N+1, k\rightarrow k+1$ with respect to the $U(N)_k$ case we propose an invariant index\footnote{In USp/SO theories the index obtained from the branes is not the Witten index, but rather a bosonic index\cite{Delmastro:2020dkz}. I thank Jaume Gomis and Zohar Komargodski for discussions on this matter.}
 \beq
I=\frac{(k+\tfrac{1}{2}N+\tfrac{3}{2})!}{(k-\tfrac{1}{2}N+\tfrac{1}{2})!(N+1)!} \label{witten-index2} \, .
\eeq
 Both the electric and magnetic branes configurations yields the same result \eqref{witten-index2}.

The phase diagram of the $USp(2N)$ theory is similar to $U(N)$ theory. For a negative fermion mass, where the duality can be trusted, the IR theory is $USp(2N)_{2(k-\frac{1}{2}(N+1))}$ TQFT. The magnetic theory, upon integrating out the massive fermion, is $USp(2(k-\frac{1}{2}N-\frac{1}{2}))_{-2N}$. The two descriptions agree thanks to level-rank duality

\beq
USp(2N)_{2(k-\frac{1}{2}(N+1))} \sim USp(2(k-\frac{1}{2}N-\frac{1}{2}))_{-2N}
\eeq
\vskip 1cm
When $\tilde N <0$ the magnetic brane configuration consists of {\it anti} D3 branes. It describes the electric theory when supersymmetry is dynamically broken. When anti D3 branes are placed on top of $O3^+$ plane the gauge group is $USp$, but the fermion transforms in the two-index {\it antisymmetric} representation \cite{Sugimoto:1999tx}, in agreement with the duality conjecture of \cite{Gomis:2017ixy}.
 
The magnetic dual is therefore
\beq 
USp(2(\frac{1}{2}N+\frac{1}{2}-k))_{2(\frac{1}{2}k +\frac{3}{4}N + \frac{1}{4})} \, + \, {\rm antisymmetric}\,\,{\rm fermion}
 \eeq
 
 Let us integrate out the fermion in order to obtain the IR theory. The level is shifted as follows:
 \beq
 \frac{1}{2}k +\frac{3}{4}N + \frac{1}{4} \rightarrow \frac{1}{2}k +\frac{3}{4}N + \frac{1}{4} \pm \frac{1}{2}(\tilde N-1)
 \eeq
 We therefore obtain for the negative mass phase a TQFT as follows 
 \beq
 USp(2(\frac{1}{2}N+\frac{1}{2}-k))_{2N} \sim USp(2N)_{(2(k-\frac{1}{2}N-\frac{1}{2}))} \, .
 \eeq
 
 The quantum intermediate phase is described by a TQFT
 \beq 
 USp(2(\frac{1}{2}N+\frac{1}{2}-k))_{2(k+\frac{1}{2}N+\frac{1}{2})} \sim  USp(2(k+\frac{1}{2}N+\frac{1}{2}))_{(2(k-\frac{1}{2}N-\frac{1}{2}))} \, .
 \eeq
\vskip 0.5cm  
  Similar to the $U(N)$ case we can use the brane configuration to introduce a magnetic' theory
\beq
USp(2(\frac{1}{2}N+\frac{1}{2}+k))_{2(\frac{1}{2}k -\frac{3}{4}N - \frac{1}{4})} \, + \, {\rm antisymmetric}\,\,{\rm fermion}
\eeq
 that covers the positive mass and the intermediate phase of the electric theory.

The low energy of the magnetic and magnetic' theories matches the low energy of the electric theory.

 \subsection{$SO(2N)$ and $SO(2N+1)$ theories}

Similar to the $USp(2N)$ theory, we can study SO theories by placing $N$ D3 branes and their mirrors on top of an $O3^-$ plane. The $O3^-$ tolarates a fractional D3 brane on it, hence we can use brane configurations to study both $SO(2N)$ theories and $SO(2N+1)$ theories. In our discussion we use the notation $SO(2N)$ with $N$ either integer or half integer. We start with $k'=k + \frac{1}{2}(N -1)$. The discussion is almost identical to the discussion about $USp(2N)$ theories. The main difference is that when the fivebranes are exchanged a D3 brane is created and hence the dual gauge group is $SO(2\tilde N)$, with $\tilde N = k'+1-N=k-\frac{1}{2}(N-1)$ and $\tilde k = k'-\frac{1}{2}(\tilde N -1)=\frac{1}{2}k + \frac{3}{4}N - \frac{1}{4}$. Therefore for $\tilde N>0$, when SUSY is not broken the duality is
\beq
 SO(2N)_{2k} \iff SO(2(k-\frac{1}{2}N+\frac{1}{2}))_{-2(\frac{1}{2}k + \frac{3}{4}N - \frac{1}{4})}
\eeq
with an index \footnote{The index calculated using branes is {\it not} the Witten index. It is rather a bosonic index, see \cite{Delmastro:2020dkz} for a comprehensive discussion.}
\beq
I=\frac{(k+\tfrac{1}{2}N-\tfrac{3}{2})!}{(k-\tfrac{1}{2}N-\tfrac{1}{2})!(N-1)!} \label{witten-index3} \, .
\eeq
The SUSY duality agrees about the low energy TQFT when the fermion mass is negative

\beq
SO(2N)_{2(k-\frac{1}{2}(N-1))} \sim SO(2(k-\frac{1}{2}N+\frac{1}{2}))_{-2N}
\eeq

\vskip 1cm

When $\tilde N$ is negative the dual magnetic theory lives on a collection of {\it anti} D3-branes placed on an $O3^-$ plane. It is $SO(2(\frac{1}{2}N-\frac{1}{2}-k))_{2(\frac{1}{2}k + \frac{3}{4}N - \frac{1}{4})}$ with a two-index {\it symmetric} fermion. The magnetic' theory is $SO(2(\frac{1}{2}N-\frac{1}{2}+k))_{2(\frac{1}{2}k - \frac{3}{4}N - \frac{1}{4})}$  with a two-index {\it symmetric} fermion. Upon integration over the symmetric fermion the magnetic and magnetic' theories describe the vacua of the electric theory in its three phases.

\section{Conclusions}

In this note we used string theory to derive new dualities as well as to recover known dualities of adjoint 3d QCD. Branes turned out to be a very useful tool to uncover a broad picture, where known dualities can be explained in terms of 3d Seiberg duality (or Giveon-Kutasov duality). 

For $k\ge \frac{1}{2}h(G)$  we found new supersymmetric dualities that hold even when a mass is given to the adjoint fermion, indicating that brane dynamics can be trusted even when supersymmetry is softly broken. For $k <  \frac{1}{2}h(G)$ we found an interesting phenomenon: Seiberg duality can be extended to the regime where supersymmetry is dynamically broken. It suggests than in 4d SQCD the Affleck-Dine-Seiberg superpotential may be computed using a dual magnetic theory. While it is long known that string theory can be used to compute the ADS superpotential \cite{Hori:1997ab}, the computation was carried out in the electric variables.  

It would interesting to generalise the results of this note to other representations (using other branes setups) and more flavours of adjoint fermions (by tilting the branes differently while preserving more supersymmetry). Another interesting direction of future work is to relate the brane dynamics to ${\cal N}=1$ supersymetric theories with flavour \cite{Bashmakov:2018wts}
.

\subsection*{Acknowledgements}

I would like to thank professors Zohar Komargodski and Vasilis Niarchos for discussions. I also thank the Simons centre for geometry and physics where this work started and to Zohar Komargodski for an inspiring discussion at the Simons centre that led to the duality idea. I am indebted to Jaume Gomis and Zohar Komargodski for a critical reading of a draft of this paper.


\newpage
\providecommand{\href}[2]{#2}

\bibliographystyle{utphys}

\end{document}